\documentclass[twocolumn,a4paper]{revtex4} % for submission
\usepackage{graphicx}
\usepackage{amssymb}
\newcommand{\etal}{{\it et al}.}
\begin{document}

\title{

Initiation of  granular surface flows in a narrow channel 
}

\author{
    Pierre Jop\footnote{Electronic address: pierre.jop@polytech.univ-mrs.fr}, Yo\"el Forterre and Olivier Pouliquen
}
\affiliation{
IUSTI, CNRS UMR 6595, Universit\'e de Provence, 5 rue Enrico Fermi,
13465 Marseille Cedex 13, France.
}
\date{\today} 

\begin{abstract}
We experimentally investigate how a long granular pile confined in a narrow channel destabilizes when it is inclined above the angle of repose. A uniform flow then develops, which is localized at the free surface. It first accelerates before reaching a steady uniform regime. During this process, an apparent erosion is observed and the thickness of the flowing layer increases.  We precisely study the evolution of the vertical velocity profile in this transient regime.  The measurements are compared with the prediction of a visco-plastic model [P. Jop, Y. Forterre and O. Pouliquen, Nature \textbf{441}, 727 (2006)].%\cite{jop06}. 
\end{abstract}    

\maketitle
\newpage % for submission only

A characteristic of dry granular materials is that they can behave like a solid or a liquid. A typical situation is obtained when an avalanche is triggered at the surface of a pile. In this case,  grains start to move at the free surface, accelerate and put into motion other grains initially static. Understanding  how the flowing part interacts with the static part  has motivated many experimental works. Different configurations have been investigated: avalanches propagating on a static layer inclined with respect to the horizontal \cite{daerr99,malloggi06,borzsonyi05}, pile collapsing on an horizontal surface \cite{lube04,lajeunesse05}, flows in rotating drums or on a pile \cite{gdrmidi04,komatsu01,bonamy02,rajchenbach00,jain02,taberlet03}. However, the dynamics observed in these experiments is complex, since the frontier between flow and no-flow evolves both in space and time. Investigating the erosion process in a uniform avalanche, where the flow/no-flow interface varies only in time, is one of the motivation of this study. 

 From a theoretical point of view, different  approaches have been proposed to describe the solid-liquid transition. A whole class of models is based on depth averaged equations and writes the mass and momentum equation for the flowing layer and the static pile  \cite{aradian02}. In this framework, an additional closure equation has to be proposed to describe the evolution of the interface \cite{bouchaud94,douady99,khakhar01}. A second approach considers the granular material like a mixture of a liquid and of a solid phase  and writes an empirical equation for the liquid phase fraction \cite{aranson02}. This approach captures some non trivial features of the transition between static and flowing regions \cite{aranson02}.
 
Recently,  it has been shown that for some configurations with sidewalls, the localization of the granular flows on top of a pile is simply related to the non uniform distribution of stresses. Due to the lateral friction,  the ratio between shear stress and normal stress decreases when going deeper in the pile. At a critical depth, it reaches the yield threshold and the material stops. Using a visco-plastic rheological model \cite{jop05,jop06}, quantitative predictions have been obtained for steady uniform flows. One can then wonder if unsteady avalanches, where erosion is observed, can be captured by the same approach. 

\begin{figure}[ht]
\begin{center}
\includegraphics[width=8.5cm]{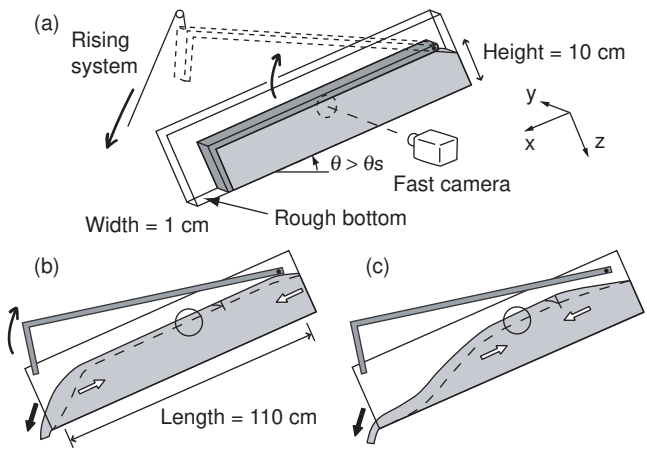}
\caption{(a) Sketch of the experimental set-up.  (b) and (c) Sketch of the pile evolution when the wedge is suddenly released. Measurements are carried out in the circled zone, before the two fronts (white arrows) reach it.  The dashed line represents the moving interface between the flowing layer and the static pile. }
\label{fig:manip}
\end{center}
\end{figure}

To study initiation of flow, we design an experimental set-up where a long pile confined in a narrow channel is suddenly destabilized above the angle of repose.  This configuration allows to create an accelerating surface flow, where the interface between flow and no-flow regions is flat and evolves in time only. The experimental set up is presented in Fig.~\ref{fig:manip}. It consists  in a long channel with glass sidewalls and a rough bottom to prevent the whole pile from flowing. We use glass beads $d=0.53\pm0.05$ mm in diameter, and density $\rho_s=2450$ kg/m$^3$. In order to study a quasi-2D system we choose a narrow channel  $W=19d$ (1 cm). The velocity of the grains measured through the glass sidewall is then  representative of the bulk behaviour \cite{jop05}. A long L-shaped  wedge closes the channel and defines a rectangular box 110 cm long and 10 cm high.  The wedge can rotate around its upper tip. At the beginning of an experiment, the channel is empty and the wedge is in the bottom position. The whole set up is inclined at the desired inclination $\theta$. The beads are then poured  below the wedge from the top.  To have reproducible initial conditions we gently tap several times on the set-up. Thanks to this set-up, a long static pile can be created at any arbitrary inclination [Fig.~\ref{fig:manip}(a)], which allows to study initiation of flow above the angle of spontaneous avalanches. 

At $t=0$, we release the heap by rapidly lifting the wedge.  The time evolution of the pile is sketched in Fig.~\ref{fig:manip}. The lower part of the pile collapses (black arrow) and creates an uphill front (white arrow). At the same time a uniform layer of grain starts to flow in the middle of the pile. Due to the downhill flow, the height of the rear of the pile decreases, leading to a downhill front (white arrow). In between these two fronts, grains accelerate but the flow properties remain uniform along the $x$-axis. All the measurements presented in the following are done in  this uniform central region (circles in Fig.~\ref{fig:manip}) before the two fronts arrive.  The measuring time allowed in this uniform state varies between $1$~s and $2.5$~s.

\begin{figure}[ht]
\begin{center}
\includegraphics[width=8cm]{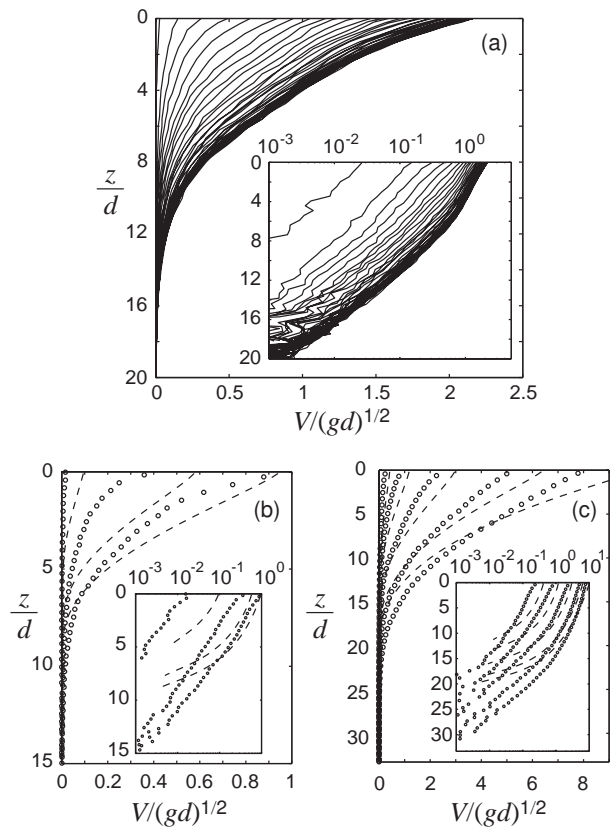}
\caption{(a) Evolution of the velocity profile at the wall $V(z)$ for  $\theta=28.0^\circ$. The time separating two curves is $\Delta t=30$ ms during 1.8 s. Inset: semi-logarithmic scale; (b) and (c) Comparison between  experiments (symbols) and numerical simulation (dashed lines)  for the velocity profiles: (b) $\theta=26.1^\circ$ at $t/\sqrt{d/g}=1.2$, $15.1$ and $166.1$ (Inset: same data in lin-log), (c) $\theta=32.15^\circ$ at $t/\sqrt{d/g}=2.3$, $7.5$, $24.2$, $77.8$ and $175$ (Inset: same data in lin-log).}
\label{fig:vdezdet}
\end{center}
\end{figure}

The motion of the grains in this central region is recorded from the side using a fast camera  (500-1500 fps). From the movies we compute the instantaneous velocity profiles using a PIV method (direct correlations).  The figure \ref{fig:vdezdet}(a) shows a typical time evolution of the velocity profile. Each profile $V(z,t)$ is obtained by averaging in the $x$-direction over the recorded window  and averaged over 10 different runs carried out in the same conditions. Here, the origin of the $z$-axis corresponds to the position of the free surface and takes into account the slight dilatancy observed during the dynamics (from 1 to 2 particle diameters).
 The first observation in Fig.~\ref{fig:vdezdet}(a)  is that the velocity profile evolves toward a steady shape, shown by the accumulation of the curves.  This saturated state corresponds to the steady uniform flow, which develops at the same angle when the beads are continuously supplied \cite{jop05}. The second observation is that the flowing layer get thicker as it accelerates, meaning that both acceleration and erosion take place simultaneously. Finally one observes that the velocity profile is first exponential, but changes shape when the layer accelerates. Still, an exponential tail is observed deep in the pile as shown in the lin-log plot [inset of Fig.~\ref{fig:vdezdet}(a)]. 

\begin{figure}[ht]
\begin{center}
\includegraphics[width=8.5cm]{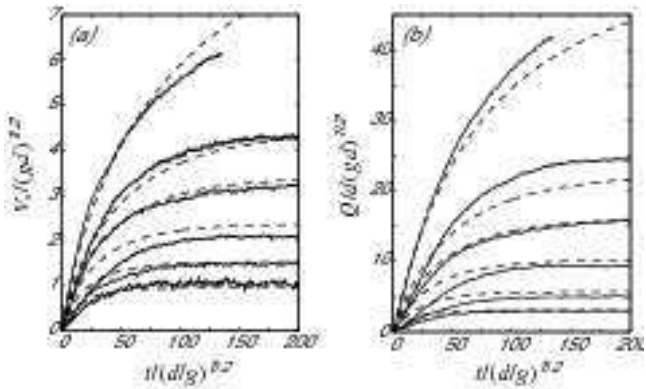}
\caption{(a) Time evolution of the free surface velocity $V_s$  for $\theta=26.1^\circ$, $27.03^\circ$, $28.0^\circ$, $28.85^\circ$, $29.5^\circ$ and $31.2^\circ$. (b) Time evolution of the flow rate $Q$; the solid lines correspond to measurements and the dashed lines to simulations.}
\label{fig:vmaxqmax}
\end{center}
\end{figure}

We have systematically  studied  the dynamics by performing experiments at different inclination angles ($26.1^\circ<\theta<32.15^\circ$). 
The time evolution of the free surface velocity $V_s(t)$ is shown in Fig.~\ref{fig:vmaxqmax}(a) for different angles (solid lines). The velocity increases and saturates when the steady regime is reached. When increasing the inclination, the acceleration is higher but it takes longer to reach the steady state. The same behaviour is observed for the flow rate $Q$  obtained by integration of the velocity profile over the depth: $Q(t)$ increases more rapidly for high angles but the steady regime establishes later [Fig.~\ref{fig:vmaxqmax}(b)].

\begin{figure}[h]
\begin{center}
\includegraphics[width=8.5cm]{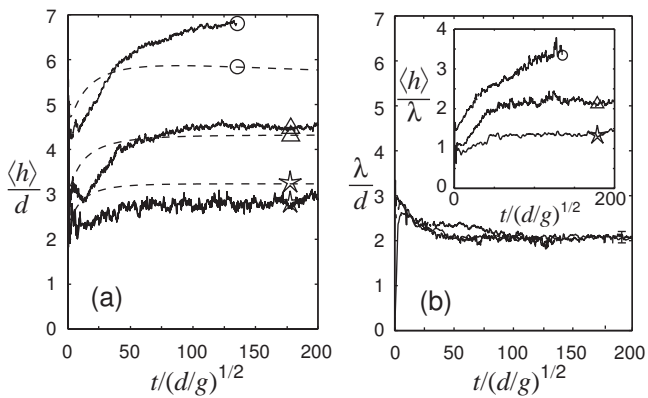}
\caption{Time evolution of the average thickness $\left < h \right >=Q/V_s$ (a) and of the  length of the exponential tail $\lambda$ (b) for $\theta=26.1^\circ$ ($\star$),  $28.0^\circ$ ($\bigtriangleup$) and $31.2^\circ$ ($\circ$). Experiments (solid lines), numerical simulations (dashed lines). Inset of (b): ratio $\left < h \right >/\lambda$.}
\label{fig:hmoy}
\end{center}
\end{figure}

To analyse the dynamics of the erosion process, we would like to define a flowing thickness. However, due to the exponential tails, there is no position corresponding to a strictly zero velocity. We have therefore define two lengthscales characterizing the velocity profile: $\lambda$ the characteristic length of the exponential tail and   $\left < h \right >(t)=Q/V_s$ the mean flow depth.  With this definition of  $\left < h \right >$,  a fully exponential profile  would give $\left < h \right >=\lambda$, whereas a fully linear profile over a thickness $H$ would give  $\left < h \right >=H/2$. Comparison between $ \left < h \right >$ and $\lambda$ then gives  information about how the profile differs from an exponential.  
The time evolution of $\left < h \right >$  and $\lambda$ are plotted in Fig.~\ref{fig:hmoy}.
The first striking observation is that the mean flow thickness $\left < h \right >$ starts from a non-zero value [Fig.~\ref{fig:hmoy}(a)]. This means that, as soon as the  avalanche starts, a finite thickness instantaneously starts flowing. Then the thickness increases and eventually reaches its steady regime value.  At the same time, the length $\lambda$ of the exponential tail slightly decreases in time,  but essentially remains constant equal to 2 particle diameters, independently of the inclination [Fig.~\ref{fig:hmoy}(b)]. This value for $\lambda$ is consistent with other measurements made in steady flows  \cite{komatsu01,bonamy02}  or in transient avalanches near the angle of repose \cite{courrechdupont05}. However, contrary to the spontaneous avalanches studied by Courrech du Pont \etal \cite{courrechdupont05}, the velocity profile in our experiments  conducted above the angle of avalanche is not always exponential. This is shown inset of Fig.~\ref{fig:hmoy}(b) where the ratio $\left < h \right >/\lambda$ is plotted for different inclinations. This ratio should be equal to 1 for a pure  exponential profile.  At low inclination, the velocity profile remains close to an exponential during the whole process, as in \cite{courrechdupont05}. However, for higher inclinations, the velocity profile differs from exponential  as soon as the flow starts  (Fig.~\ref{fig:vdezdet}(c) and the ratio $\left < h \right >/\lambda$  increases during the flow acceleration.

Our measurements then show that the velocity profile follows a non trivial dynamics where both accelerations and erosion are present. In the following, we try to compare these measurements with the prediction of the constitutive law proposed  in \cite{jop06}. We know that the rheology quantitatively described the assymptotic steady regime \cite{jop05}, and we would like to test if it can also capture the flow initiation.  
The rheology assumes that the granular material behaves like an incompressible fluid of density $\rho_s \Phi$ ($\Phi$ being the volume fraction equal to 0.6) with a visco-plastic constitutive law.  The relation between the stress tensor  $\sigma_{ij}$  and the shear rate tensor $\dot \gamma_{ij}$ is given by:
\begin{equation}
\sigma_{ij}=-P\delta_{ij}+\tau_{ij} { \rm  \hspace{7 mm} with \hspace{7 mm}} 
\tau_{ij}=\frac{\mu(I)P}{|\dot{\gamma}|}\dot{\gamma}_{ij},   \label{eq:rheol}
\end{equation}
where   $P$ represents an isotropic pressure, $|\dot{\gamma}|$ represents the second invariant of the shear rate tensor $|\dot{\gamma}|=\sqrt{\frac{1}{2}\dot{\gamma}_{ij}\dot{\gamma}_{ij}}$  and $\mu(I)$ is the friction coefficien \cite{dacruz05,gdrmidi04,forterre03}, which depends on the inertial number $I=\frac{|\dot{\gamma}|d}{\sqrt{P/\rho_s}}$ as follows:
  \begin{equation}
\mu(I)=\mu_s + \frac{\mu_2-\mu_s}{I_0/I+1}.
\label{mudei}
\end{equation}

In the configuration studied here, the velocity  is  parallel to the free surface  and depends on $z$ and $y$, $\vec{v}=V(y,z,t) \vec{e_x}$, and the pressure is given by  $P=\rho_s \Phi g z\cos \theta$. 
To simulate our experiments, we then have to solve the  Cauchy equation: 
$$\rho_s \Phi \partial V /\partial t=\rho_s \Phi g \sin \theta+\partial \sigma_{xz}/\partial z + \partial \sigma_{xy}/\partial y,$$ 
with the following boundary conditions: free-stress at the free surface,  no-slip condition at the bottom and a Coulomb friction on sidewalls.  To quantitatively compare the prediction with the experiments, we choose $\mu_s=\tan(20.9^\circ)$, $\mu_2=\tan(32.76^\circ)$ and $I_0=0.279$ as used in \cite{jop05}. We also need the value of the wall friction coefficient $\mu_w$. We choose $\mu_w=\tan(13.1^\circ)$, which gives the best agreement for the mean flow rate in the steady uniform regimes based on data given in \cite{jop05}.  With this choice, no free parameter remains to quantitatively predict the dynamics of our avalanches.  

\begin{figure}[h]
\begin{center}
\includegraphics[width=7cm]{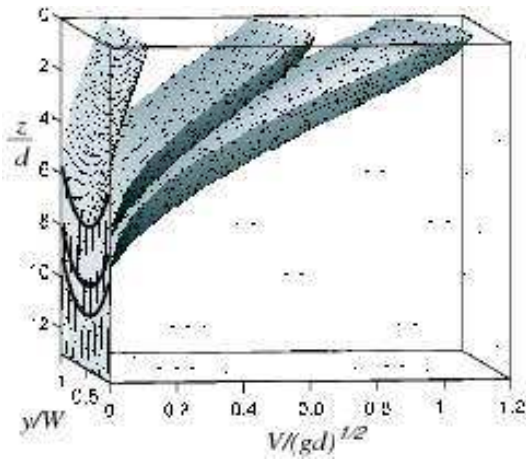}
\caption{Velocity profiles from the simulation for $\theta=26.1^\circ$ at different times $t/\sqrt{d/g}=1.5$, $13.4$ and $200$ (stationary regime). The solid lines represent the frontier between the flowing layer and the static bottom for the three profiles.}
\label{fig:exsim}
\end{center}
\end{figure}
The simulation is carried out by imposing  a zero initial velocity and by letting  the system evolve under  gravity (Fig.~\ref{fig:exsim}).  At $t=0$ a finite layer of material starts flowing. This layer then accelerates and the flow thickness increases in qualitative agreement with the experiments. In order to quantitatively compare the simulation with the experiments, we extract from the simulation the velocity profile at the sidewalls and compare with the experimental data. The prediction are plotted in Figs.~\ref{fig:vdezdet}(b) and \ref{fig:vdezdet}(c) (dotted lines) for  $\theta=26.1^\circ$ and $\theta=32.15^\circ$. We see that the agreement for the evolution of the velocity profile is good at  high inclination, but becomes less accurate at low inclination. In the model the velocity goes to zero at a finite depth, whereas in the experiments the profiles are followed by an exponential tail. At low angles the experimental profiles are mainly exponential and the model does not capture the correct shape. 

In order to more precisely check if the dynamics is correctly predicted by the model,  we extract from the simulations  the same quantities computed in the experiments, namely the free surface velocity at the wall $V_s(t)$, the  flow rate at the wall $Q(t)$ and the mean thickness $\left < h \right >(t)$. The results are the dotted lines in Fig.~\ref{fig:vmaxqmax} and  Fig.~\ref{fig:hmoy}.  We find a good agreement for the free surface velocity and the flow rate, the time evolution being quantitatively captured by the model for all inclinations.  The agreement is less accurate for the mean thickness $\left < h \right >(t)$. As shown in Fig.~\ref{fig:hmoy}, the model predicts that $\left < h \right >(t)$ is initially non zero and increases with time in a similar way the experiments do. However, quantitatively the prediction is not very accurate. The major reason for this discrepancy is that the parameter $\left < h \right >(t)$ is sensitive to the exponential creeping tail, which is not captured by the model.  

In conclusion, we have shown that  in a narrow channel configuration, for which wall effects are predominant, the observed erosion process is captured by a simple visco-plastic model. The inertia and the existence of a threshold are sufficient to create an "erosion-like" phenomenon. The frontier between flow and no flow regions is simply driven by the time evolution of the internal stresses. The model quantitatively predicts the time dynamics  showing the relevance of the proposed rheology for unsteady flows.  On the other hand, serious limits exist. First, the model fails in precisely predicting the velocity profile shapes, mainly because the creeping exponential tail is not captured. The discrepancy is even more important at low inclinations, close to the angle of repose. 
The second limit concerns the lack of hysteresis in the model. Our study is restricted to avalanches triggered above the angle of repose. To capture natural avalanches observed for example when  slowly inclining a pile, or to describe intermittent flow regime, hysteresis has to be taken into account in the model. These limits shows that further developments are needed for a more complex rheology, which could  capture both the quasi-static limit and the hysteretic nature of granular flow threshold.

\end{document}